# Challenging the challenge: handling data in the Gigabit/s range

T.Anticic, F.Carena, W.Carena, R.Divià, D.Favretto, J.C.Marin, A.K.Mohanty, B.Polichtchouk, F.Rademakers, K.Schossmaier, P.Vande Vyvre, A.Vascotto for the ALICE collaboration
*CERN, Geneva, CH-1211, Switzerland*
J.P.Baud, M.Collignon, F.Collin, B.Couturier, J.D.Durand, J.M.Jouanigot, B.Panzer-Steidel, H.Renshall, M.Schulz
*CERN, Geneva, CH-1211, Switzerland*

The ALICE experiment at CERN will propose unprecedented requirements for event building and data recording. New technologies will be adopted as well as ad-hoc frameworks, from the acquisition of experimental data up to the transfer onto permanent media and its later access. These issues justify a careful, in-depth planning and preparation. The ALICE Data Challenge is a very important step of this development process where simulated detector data is moved from dummy data sources up to the recording media using processing elements and data-paths as realistic as possible. We will review herein the current status of past, present and future ALICE Data Challenges, with particular reference to the sessions held in 2002 when – for the first time – streams worth one week of ALICE data were recorded onto tape media at sustained rates exceeding 300 MB/s.

## 1. INTRODUCTION

All the experiments installed at the LHC collider at CERN announced out of the usual requirements. Data streams of unprecedented rates and volumes will be established between detectors, computer farms (online and offline) and mass storage systems. A reliable and effective cooperation will be expected from several components, hardware and software, in-house, public domain and commercial. Final objectives: satisfy the initial requirements, allow subsequent expansions and ensure the desired performance with the maximum reliability. The ALICE experiment [1], with its very high-volume data streams, makes no exception to the rule. To guarantee the viability and reliability of the ALICE Data Acquisition, Data Handling and Permanent Data Storage systems, periodic tests are held in collaboration with the ALICE Online and Offline teams together with the CERN central services (Permanent Data Storage, Operating System deployment and support, centralized data repository and distribution, networking) – the so called ALICE Data Challenges. In this paper we review the past experiences of the ALICE Data Challenges, the achieved milestones during the current production period and the future plans.

## 2. ALICE AND THE DATA CHALLENGES

The main purpose of the ALICE experiment is to study strongly interacting matter under conditions of extreme temperature and density using beams of heavy ions, such as those of lead. The particles in the beams will collide thousands of times per second and each collision will generate an event containing up to thousands of charged particles. Thus, every second, the characteristics of thousands of particles will have to be recorded. A central ALICE event, with lead beams, contains approximately two orders of magnitude more data than ATLAS or CMS events with a proton beam.

### 2.1. ALICE running parameters

The data stream of ALICE will be made of several types of events, each with its own unique signature. Central and Minimum Bias events will be acquired with a relatively low rate – around 10 events per second – for a high data volume of 10 to 40 Megabytes per event. Dielectron events – where a partial readout scheme (channels with uninteresting data will not be read out) will be used – should produce a higher rate data stream (about 100 events per second) for a smaller data volume (1 to 4 Megabytes per event). To complete the list, there will be events with Dimuon trigger signatures, which are expected to generate some 1500 events per second for a data size of 200 to 750 Kilobytes per event. In summary, there will be three types of event, each one contributing to about one third of the final data volume going through the ALICE Data Acquisition system. Adjustments made in real time to the behavior of the trigger system will avoid the starvation of some classes of events due to congested data paths between the detectors and the Data Acquisition system. All these factors will create a global data stream with a rather complex structure.

The ALICE detector is expected to be ready to run with the above parameters for one month a year, 24 hours a day and the maximum achievable availability. The data flow between the Data Acquisition system and the Permanent Data Storage will be limited to a throughput of 1.25 Gigabytes per second, for a grand total of 1 Petabytes produced during the lead beam period. Another half a Petabyte will be created each year during the proton beam period. This will result in a yearly production of 1.5 Petabytes of data to be recorded, catalogued, labeled and made available for later processing. Higher data volumes are expected inside the Data Acquisition system, where several data compression and filter stages will take place at various places.

In a summary, several challenges are proposed to the ALICE collaboration:

1. Handling of low-rate, high-volume events.
2. Handling of high-rate, low-volume events.
3. Online Filtering and Data Compression stages.
4. Handling of high-volume stream to Permanent Data Storage
5. Effectiveness, reliability and availability for all of the points above.

6. Indexing and access to the experimental data for distribution and processing functions (filtering, reconstruction, analysis) during and after data collection and recording.

To perform all these tasks, many packages have to be developed, debugged and validated. Commercial and Open-Source products have to be evaluated, installed, configured and tuned. Common software has to be agreed upon, developed and integrated. Hardware must be built or purchased, evaluated, assembled, validated and put in operation. This is clearly a highly challenging task, spanning over several years and covering many disciplines. That is where the ALICE Data Challenges play a vital role in the preparation process.

## 2.2. The ALICE Data Acquisition System architecture

The ALICE Data Acquisition System (DAQ) architecture will be based on a data-driven approach. Under the control of a three-level trigger system, the Front End Electronics (FEEs) – located as closed as possible to the detectors – will readout, format and validate the event raw data at a "local" level (ranging from a complete detector to a sector or a sub-sector of the same). All accepted events will be shipped via a custom-designed point-to-point optical link called Detector Data Link (DDL) to a Local Data Concentrator (LDC), a commodity PC located a few hundred meters away from the interaction point. The LDC will validate the event, eventually perform local event building (for LDCs with multiple incoming DDLs), run data compression and other data analysis functions and finally move the raw data to the event builder, running on a Global Data Collector (GDC). On the GDC – again based on a commodity PC – the full event will be assembled in the host memory and will be made available for further processing stages and for recording. The Permanent Data Storage system (PDS) will perform data recording functions and will provide access to the event data and catalogues for all successive analysis stages.

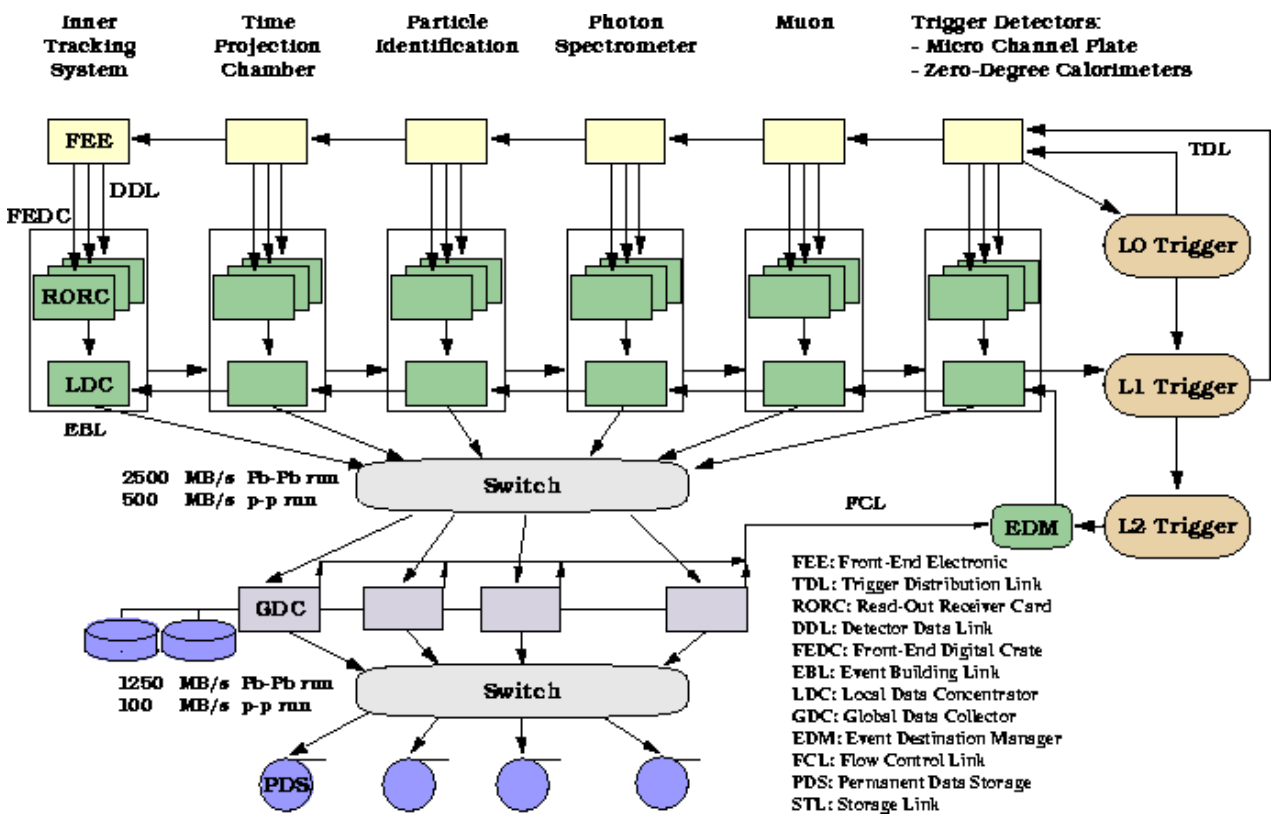

Figure 1: ALICE Data Acquisition System architecture.

Looking at the main data flow, while the FEEs and the DDLs will be based on custom-designed components, the LDCs, GDCs and the PDS, together with their associated networks, will be built using commodity hardware. The actual technologies to be used for the final ALICE setup will be decided as late as possible and will eventually be upgraded during the lifetime of the experiment. A staged installation strategy has been decided in order to add progressively new material during the first two years of data taking, as soon as higher rates will be required. This will allow a considerable reduction in the overall expenses as well as a more efficient final system.

## 2.3. Requirements and planning for the ALICE Data Challenges

Target of the ALICE Data Challenge is to put together all the elements available at a given moment in time and to create a chain as complete as possible, from the data sources (simulated at different levels) to the Permanent Data Storage. State-of-the-art technologies are for the first time integrated into a single chain to evaluate the individual and global behaviors. Components are installed and tuned to match the relations with the rest of the chain. By achieving – year after year – more challenging targets, we expect to setup, right before LHC startup, a system up to the requirements of the ALICE experiment.

In Figure 2 is the planning of the ALICE Data Challenges as function of the targeted data rates through and recorded by the Data Acquisition system in agreement with the deployment planning of the LCG[1] testbed, where the Challenges do and are expected to take place.

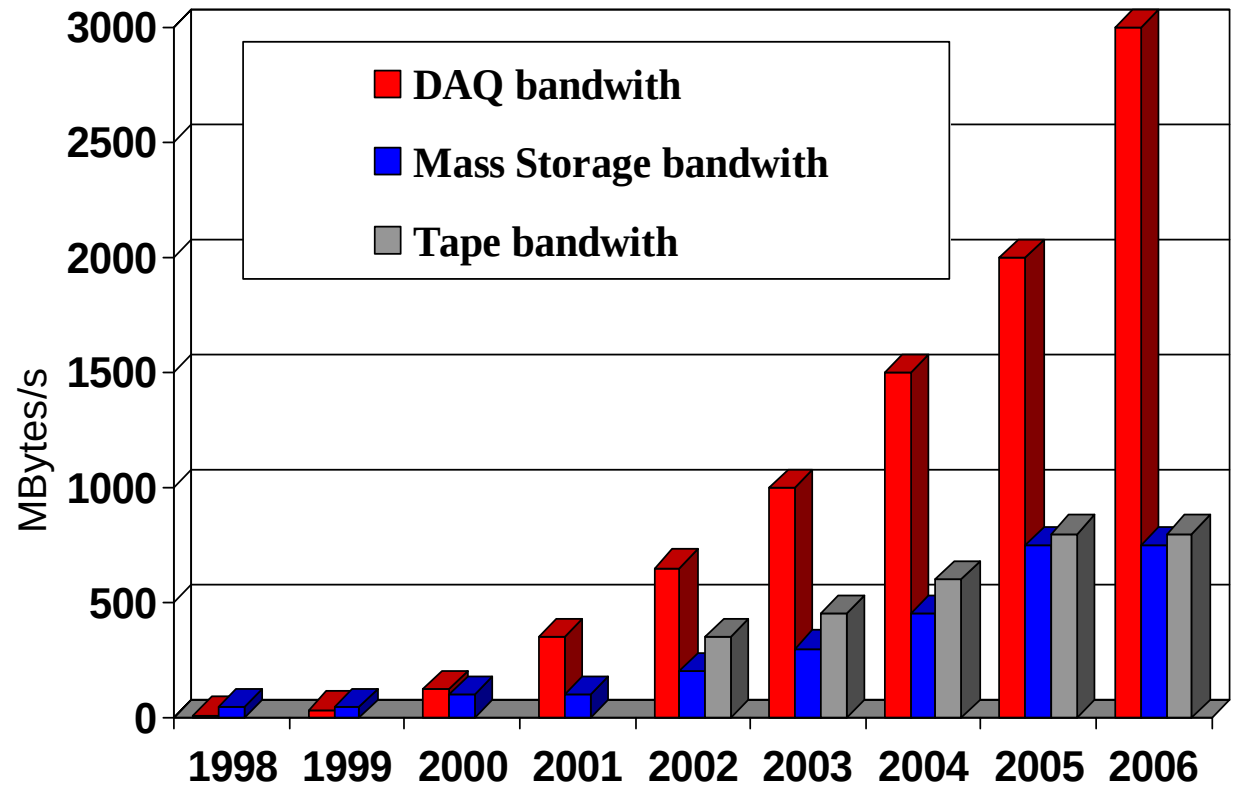

Figure 2: ALICE Data Challenges bandwith planning.

As we can see from Figure 2, the target is to progressively increase the data rates until something as close as possible to the final ALICE requirements – in agreement with the available hardware resources allocated to the exercise – shall be met. We expect this will happen at last one year before LHC startup.

[1] LCG stands for LHC Computing Grid Project, for more information see http://lcg.web.cern.ch/LCG

Seen the unprecedented quantity of data to be stored in the Permanent Data Storage, we may witness problems of scalability, data query, data retrieval and concurrent data access. Therefore, the volume of recorded data during the Data Challenges needs also to be planned based upon the requirements of the ALICE experiment. A set of milestones has been established concerning the data volumes. These milestones can be seen in Figure 3.

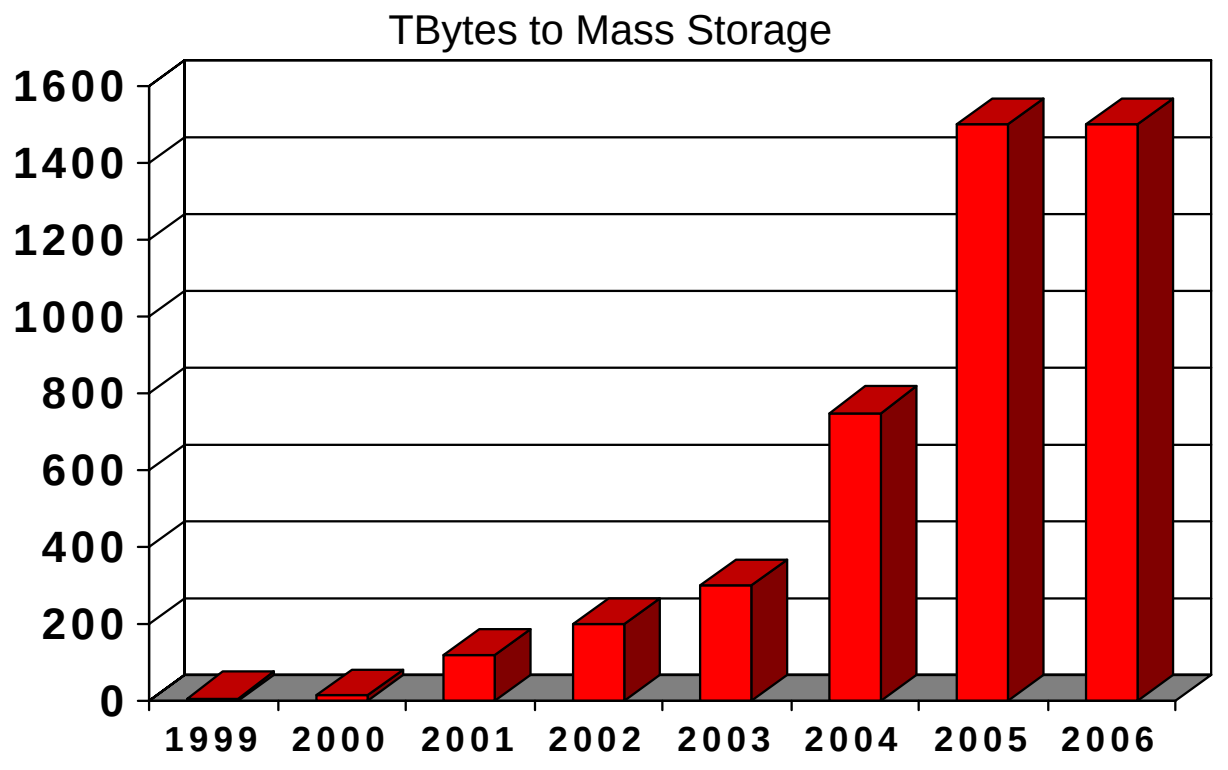

Figure 3: ALICE Data Challenges recorded data planning.

Similarly to the bandwith planning, the milestones of the recorded data planning will reach the expected requirements from the ALICE collaboration in time for the LHC startup.

### 2.4. ALICE Data Challenges: past and present

The first ALICE Data Challenge took place in 1998. In that year, several novel technologies and tools were used for the very first time for data acquisition systems at CERN – within ALICE test beams and elsewhere: Fast Ethernet links, high bandwidth network backbones, new Permanent Data Storage media and equipment, Unix-based (within ALICE: Solaris and AIX) data acquisition systems, early prototypes for the ALICE Data Acquisition and test beam environment (DATE) package, specialized storage systems such as HPSS and in-house packages and storage access libraries (direct ancestors to the current CERN advanced storage manager – CASTOR). The raise of concerns about possible interoperability and functional problems justified the setting and operation of a dummy data acquisition chain during periods of reduced activity at CERN.

The exercise was considered to be very fruitful. Several problems were spotted, eventually solved, and a considerable work of debugging and tuning took place in a relatively "relaxed" environment, where reliability and availability of the complete system was somehow less critical than in an equivalent production setup. The decision was therefore taken to periodically repeat this activity [2].

Four Data Challenges have been held so far. Each of them replaced existing components with more recent versions and introduced new elements in the chain: Operating Systems such as Linux, new – for the Data Acquisition culture – processors (Intel Pentium), Permanent Data Storage systems (CASTOR) and hardware architectures (IDE-based disk and tape servers), tape technologies (linear tape technologies such as the STK 9940 series) and networking solutions (Ethernet trunking, Gigabit and 10 Gigabit technologies).

## 3. THE ALICE DATA CHALLENGE IV

The ALICE Data Challenge IV took place between June and December 2002. Equipment from the ALICE DAQ group and from the LCG testbed was used throughout the various phases of the exercise. New versions of already used packages (DATE, CASTOR) went for the first time in operation.

### 3.1. Planned objectives

The following objectives were proposed for the ALICE Data Challenge IV:

1. Scalability test for the Data Acquisition system to control and handle hundred of nodes.
2. Data transfer inside the Data Acquisition system at 650 MB/s minimum sustained throughput for a few hours.
3. Data recording to Permanent Data Storage at 200 MB/s minimum sustained throughput for seven consecutive days.
4. 200 TB of data being recorded to Permanent Data Storage.

### 3.2. The components

As usual for the ALICE Data Challenges, the target was to use the latest available hardware and software components, namely:

1. Network technologies: trunking, backbone switching, Gigabit and 10 Gigabit Ethernet.
2. Commodity hardware: hosts, network interface cards, tape units and tape robots.
3. ALICE Data Acquisition system (DATE [3] v4) with its services (readout, monitoring, configuration, control, event building, event recording, messaging system).
4. ALICE fabric monitoring software (AFFAIR [4]) to assess the behavior of the components of the Data Acquisition system and the interface to the Permanent Data Storage.
5. ALICE Offline software: objectification of raw data, handling of event objects, recording and interfacing to the Permanent Storage System.
6. CERN Advanced Storage Manager (CASTOR [5]) – deployed on CPU servers, DISK servers and TAPE servers – for Permanent Data Storage functions.
7. Operating system (Linux) with its kernel, system and user libraries, drivers, file systems (local and networked), network daemons

(standard and custom designed) plus all CERN-specific add-ons and configurations.

All the above components were – in one-way or another – deployed for the first time within an ALICE Data Challenge.

## 3.3. Hardware setup

The hardware setup used for the ALICE Data Challenge IV can be split in four partitions:

1. DAQ emulation and support.
2. CASTOR support.
3. Networking.
4. Infrastructure.

The ALICE DAQ group and the ADC, CS and DS groups of the CERN/IT division jointly provided the environment to support the ALICE Data Challenge IV. Two computer farms – both located on the CERN main site but quite far apart – were effectively seen as one big unit thanks to the excellent CERN network backbone and to the uniform deployment of Operating Systems, packages and environments provided by the CERN Linux, ASIS[2] and AFS teams. Software could be shared without problems on all the machines with no need for explicit copying or recompilation processes. NFS guaranteed the required effectiveness, reliability and reconfiguration capabilities requested by the exercise – all issues about its scaling capabilities being dropped as more and more machines were flawlessly added to the test setup.

The hosts used for the test were all SMP-based. The main production periods took place on the LCG testbed, based on boards equipped with dual Pentium III running at ~1 GHz, an architecture that matches well the planning from the ALICE DAQ software for LDCs and GDCs. To evaluate the behavior of systems equipped with more CPUs, some tests were performed on specialized servers belonging to the ALICE DAQ group test environment.

Such a challenging exercise required an out-of-the-ordinary network setup. The test made use of CERN backbone resources as well as of dedicated equipment. Core of the network architecture were two high-bandwith switches – based on Gigabit Ethernet technologies – directly linked to a set of satellite switches each handling a group of up to twelve hosts.

Figure 4 shows the network setup for the LCG testbed, where the raw DATE performance and the Data Challenge production periods took place. In the centre of the diagram are the two central switches – Extreme Networks Summit 7i with 32 Gigabit Ethernet ports each – while LDCs, GDCs and DISK servers were connected to smaller 3COM 4900 switches (16 Gigabit Ethernet ports each). The standard CERN backbone, supported by Enterasys SSR8600 routers (28 Gigabit Ethernet ports) guaranteed the liaison with the TAPE servers and with the rest of CERN, including the ALICE DAQ lab and the workstation used for operation and control. Several trunks were deployed between switches whenever the requested bandwith exceeded the capacity of a single link.

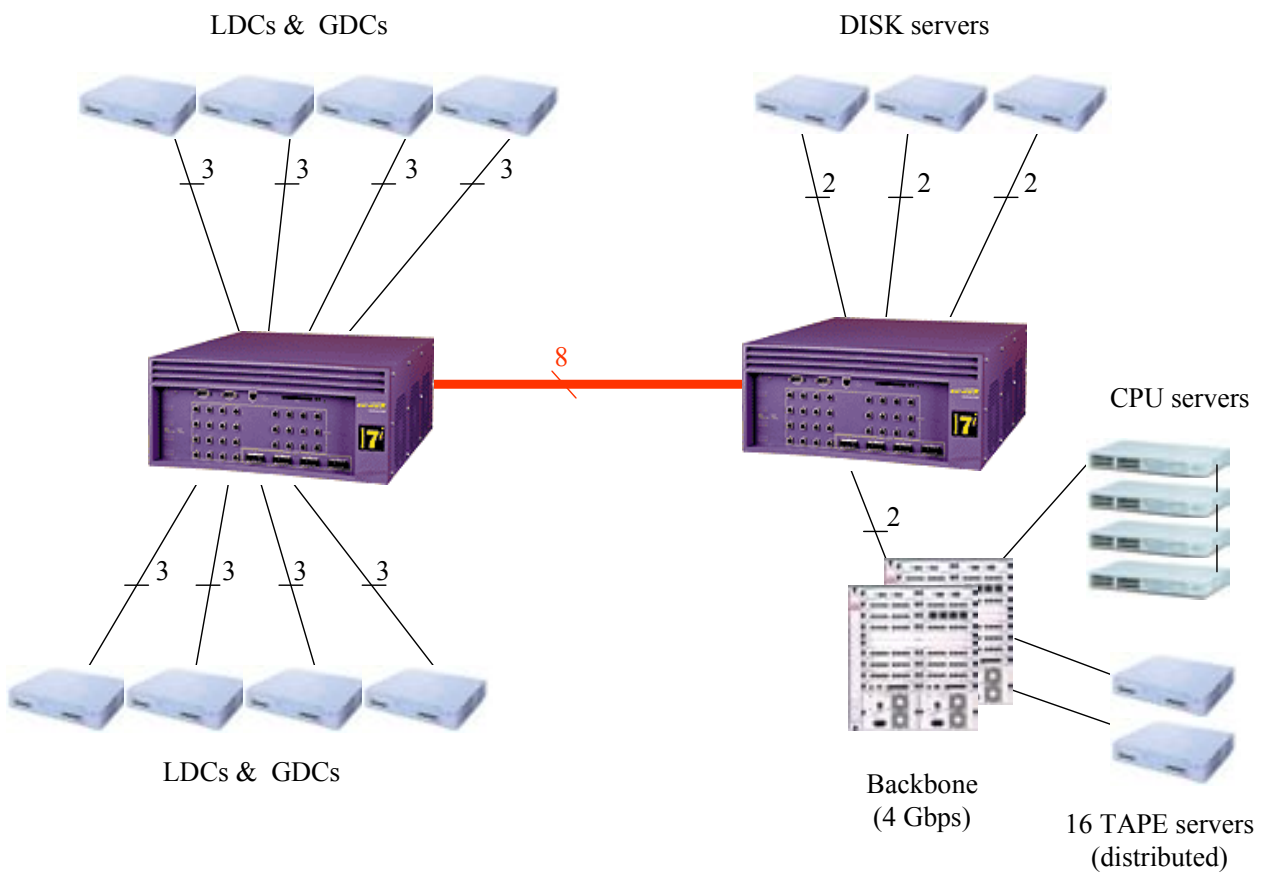

Figure 4: LCG testbed network setup.

During the exercise, equivalent 10 Gigabit Ethernet devices, in evaluation at CERN, have also been tested in network architectures similar to the one described above.

## 3.4. Software components

For the deployment of the ALICE Data Challenge IV we took standard, out-of-the-box components, integrated by ad-hoc configuration and installation tools.

The three CERN-developed packages that played a key role in the Challenge were DATE, CASTOR and ROOT.

DATE – the name stands for Data Acquisition and Test Environment – is the framework of the ALICE Data Acquisition systems, also used for R&D and test beams support. The release used – identified as version 4 – introduced several novelties, including new run control, data recording and event building packages. More scalable than its predecessors, DATE version 4 made a better use of the system resources for large-scale setups – such as the one deployed during the ALICE Data Challenge IV. Included in DATE were also a Configurable LDC Emulator (COLE) – capable of producing a ALICE-like data traffic pattern – and A Fine Fabric and Application Information Recorder (AFFAIR) package providing the required run-time global and local behavioral monitoring capabilities throughout the whole Data Acquisition system. All the information collected with AFFAIR was promptly published on WWW for immediate feedback. Several of the graphs presented in this paper have been extracted from the pages published via AFFAIR.

The CERN Advanced Storage Manager (CASTOR V1.4.1.7) package played a key role for the support of the data created by DATE during the exercise. CASTOR provided a common access library and a set of transparent migration engines to a unique name space, integrating several mass storage systems placed at different levels and support medias. All this under the pressure of hundreds of machines producing data at their maximum speed for a period of several days. Monitoring tools and public status

[2] ASIS stands for "The Application Software Installation Server", more information is available at the Web address http://asis.web.cern.ch/asis/

pages were provided to configure, operate and control the behavior of the system in real time.

A common Operating System was used throughout the test setup. Linux RedHat 7.2, kernels 2.2 and 2.4, as provided by the CERN Linux support team, was deployed on all the machines. The standard installation procedure available at CERN was used. A configuration of the system parameters was performed to appropriately size common resources such as IPC shared memory block size and TCP/IP socket size, to import the central distribution repository and to install the required network services. This required no changes in the kernel itself and could be done either on the fly or during the boot procedure of the Operating System. AFS was installed on all the machines but was not used at runtime: its role was to support the ASIS environment, for distribution of system images and CERN-wide packages. A special ALICE–developed driver for the support of shared pinned memory was installed on some of the machines used to run peer-to-peer tests.

The ALICE collaboration makes intensive use of the ROOT framework. To support the DATE built-in packages based on ROOT, including the ALICE Mock Data Challenge objectifier (ALIMDC), ROOT V3.03 was installed and distributed via NFS. Run-time libraries were also distributed via NFS and automatically loaded by DATE whenever this was required.

## 3.5. Peer to peer tests

Several peer to peer tests were made to evaluate the behavior of the key network components, namely the DATE recording library, the architecture of the DATE event builder data receiving engine and the various system libraries required by the data recording process. The tests took place in the ALICE DAQ test setup. As we did not need a complete Data Acquisition system, only a subset of the Data Acquisition components was used for this test. The chain included a minimal skeleton and the DATE recording library on one side and a data sink on the receiving end, using the same architectures as for the DATE recorder and event builder packages. Both sides of the test were extended to allow precise measurements for key system and network resources. The outcomes of these tests were very encouraging and more than validated the effectiveness of all the above components.

One of the issues to be analyzed during the peer-to-peer tests was the transfer speed and relative load on the sending and receiving CPUs. The results gave different results on dual-CPU and quad-CPU machines.

On dual-CPU hosts, whose architecture well matches the requirements for the ALICE LDCs and the low-cost ALICE GDCs, performances of up to 83 MB/s were reached with a usage of 1.5 %/CPU/Megabyte on the LDCs and 2.1 %/CPU/Megabyte on the GDCs. The machines used for this test were equipped with two Pentium III CPUs running at 1 GHz, Linux Kernel 2.4.18 and NetGear GA620 NICs with acenic driver. Detailed figures from one of the tests – where the correlation between socket sizes and throughputs for fixed-size events – are reported in Figure 5.

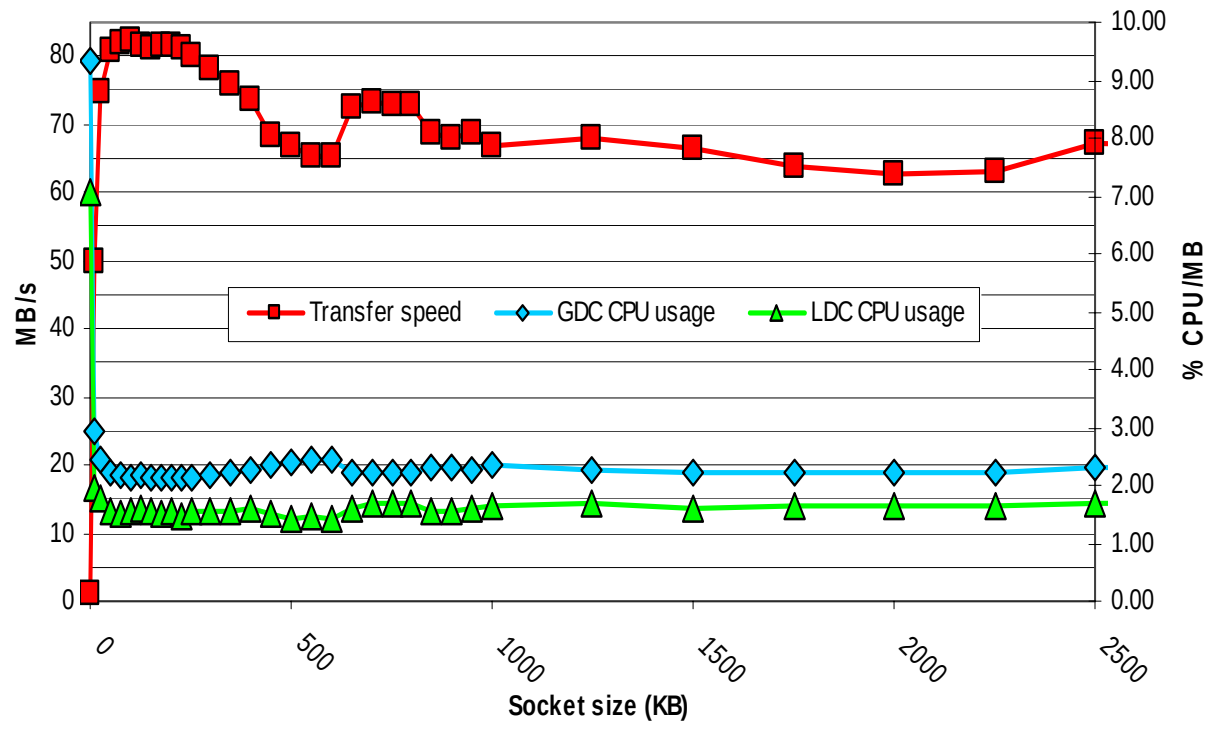

Figure 5: Peer to peer tests on dual-CPU hosts.

The same test, run on quad-CPU hosts, returned higher network performance for similar CPU usage, effect due to the large total CPU capacity and to the fact that the Linux kernel has proved to be able to make use of more than one CPU for its internal tasks. The platforms used for this test were HP Netservers with 4 Xeon CPUs running at 700 MHz, Linux kernel 2.4.19 and 3COM 996 as NICS with tg3 driver. The top performance was of 110 MB/s (very close to the Gigabit Ethernet wire speed) for 1.9 %/CPU/Megabyte on the event builders and 1.4 %/CPU/Megabyte on the data producers. Details on one of the quad-CPU tests – correlating the event size to the achieved throughput – are reported in Figure 6.

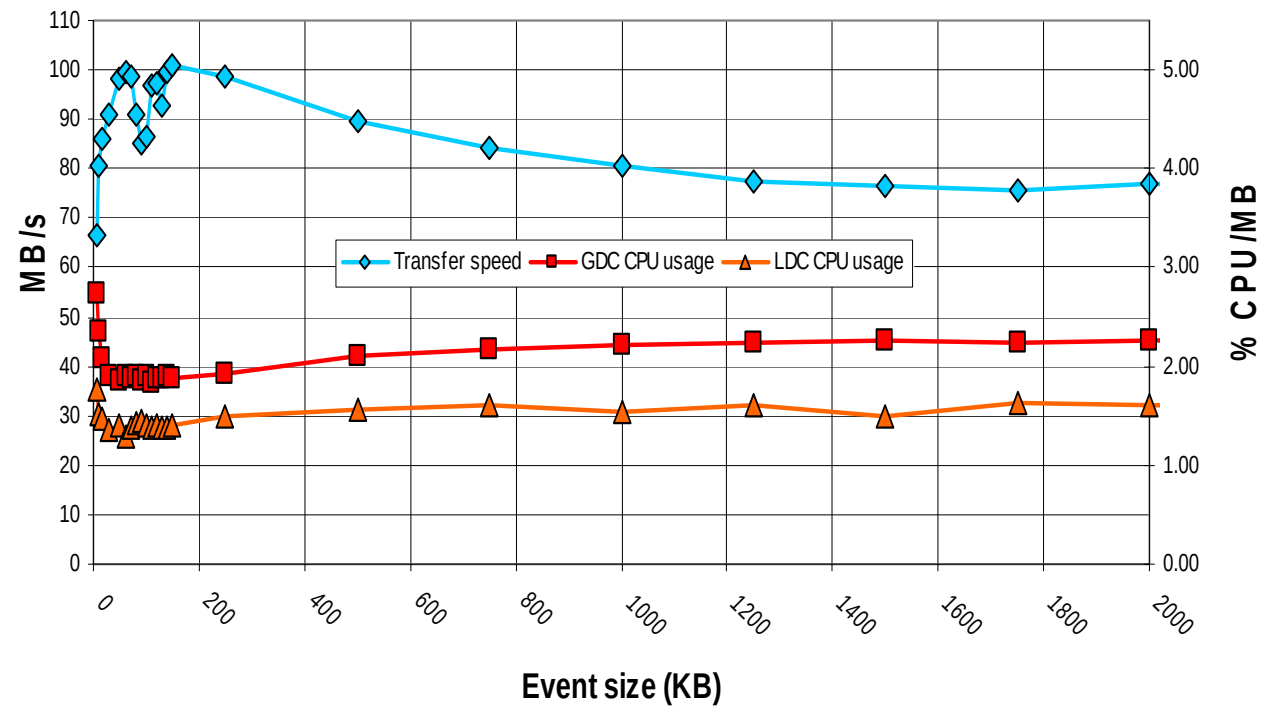

Figure 6: Peer to peer tests on quad-CPU hosts.

During the peer-to-peer tests we have also done some measurements on the possible correlation between runtime parameters (Operating System level and user code level) and data traffic behavior. The most surprising conclusion we have achieved was that the old rule of thumb "the bigger the socket, the better the performance" seems not be any longer true. We have witnessed a degradation of the throughput whenever the socket exceeded a certain size (variable with the architecture, the payload and other user code parameters). Furthermore the size of the user DATE raw data buffer proved to have a significant effect on the overall performance, following a similar pattern as in the case of the socket size.

### 3.6. Scalability tests

Target of the scalability tests was to correlate the stability and usability of the Data Acquisition system as a function of the number of LDCs and GDCs, on a scale as close as possible to the one of the final ALICE Data Acquisition system.

For this exercise, the accent was placed on the scalability of all the components. Data transfer had to give its proof of feasibility and nothing more. Key elements to validate were: the state machines controlling the whole system, the operator user interface, the communication libraries, the system I/O libraries, the usage of the system I/O libraries, the distribution system for images, libraries and configuration parameters, the information and error logging facilities and the behavior of the system as a whole.

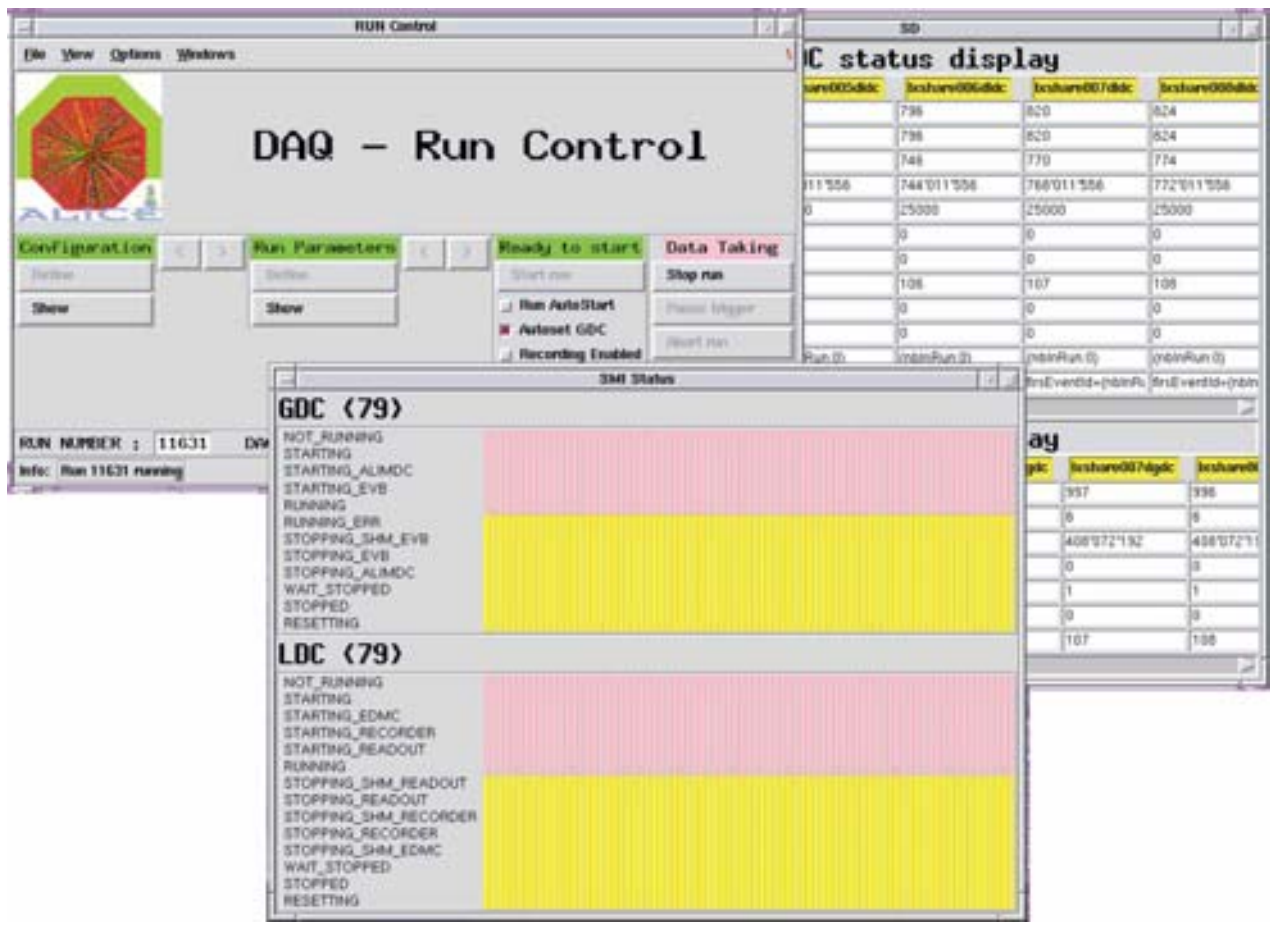

Figure 7: Status & control window during scalability tests.

The system reacted very well. No hard limitations were found in any of the components. As shown in Figure 7, a maximum configuration of 79 LDCs and 79 GDCs – running on 79 dual-role PCs – could be controlled with very acceptable latencies (a delay of ~15 seconds was measured during the start of run phase). The Operating System, together with the communication, run-time and graphics library accomplished their tasks without problems. The Operator graphic interface could effectively describe the evolution of the Data Acquisition System startup procedure even on such a large set of nodes with an excellent quality of visual hints and diagnostic information.

### 3.7. Staging of the production period

The ALICE Data Challenge IV was planned on a production chain based on the following components:

1. The ALICE LDC emulator COLE, to create a data stream according to the test requirements.
2. The ALICE Data Acquisition and Test Environment DATE, to acquire, build and record the ALICE raw data.
3. The ALICE Mock Data Challenge objectifier, to create ALICE data objects and to write them in ROOT format.
4. The CERN Advanced STORage manager CASTOR, to handle the Permanent Data Storage access (read, write, migration, operation, monitoring).
5. A Fine Fabric and Application Information Recorder AFFAIR, a monitoring system for LDCs and GDCs.

The chain was setup little by little, not necessarily in the order given above. At times, intermediate components were tested in isolation. In other tests, small parts of the chain were put in operation, to evaluate the interdependence of the various components. The outcomes of this type of exercises proved to be a valuable aid for tuning, debugging and validation of the test environment.

The LDC emulator was setup to produce one of two traffic patterns: either the so-called "flat traffic", where all the LDCs would create an identical event, or an "ALICE-like traffic", where a model of the forecasted ALICE raw data was followed. Tests were run with both types of traffic with different results.

Using flat data traffic, scalability tests were run up to the recording stage of DATE. Here we have observed a good behavior of the whole system excepted for an undesired phenomenon at the output of some of the network switches. As can be seen in Figure 4 above, the LDCs and the GDCs were all connected to one of the "satellite" 3COM 4900 switches and – via a triple Gigabit Ethernet trunk – to one of the two central Summit 7i. We therefore expected the outgoing data traffic from each of the 3COM switches to reach throughputs of the order of three times the throughput of a single Gigabit Ethernet link. Unfortunately this was not what we have measured.

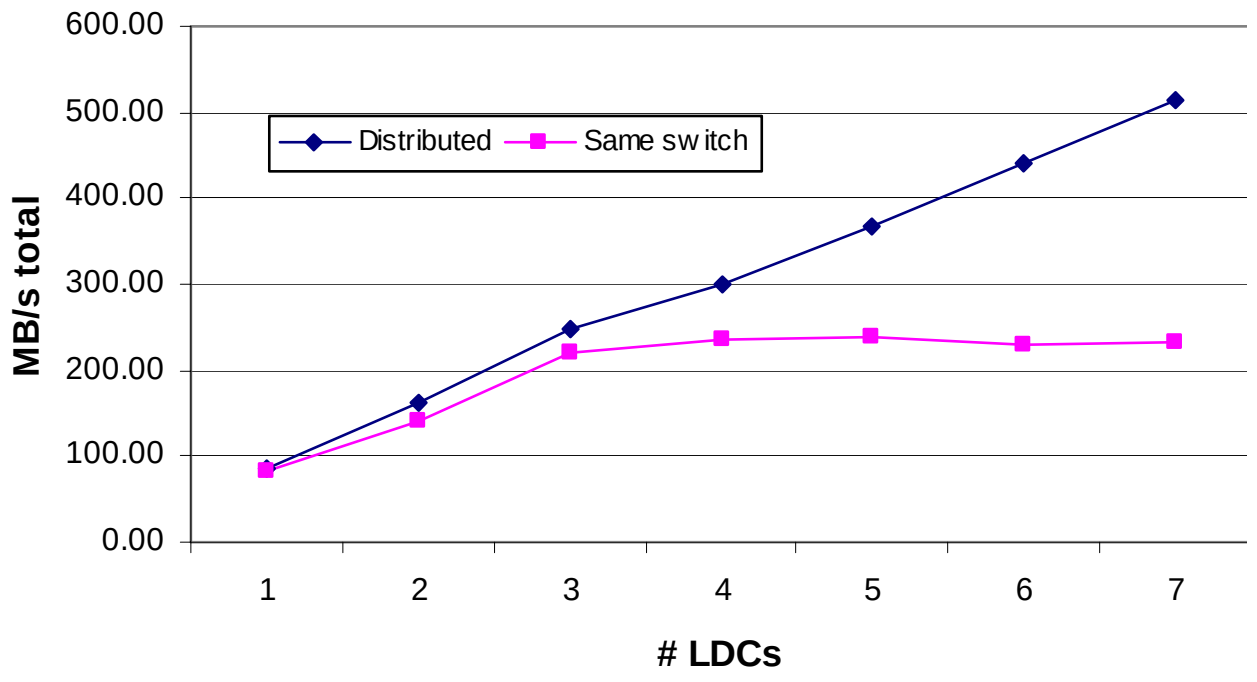

Figure 8: Scalability test of outgoing data trunks.

As we can see in Figure 8, by distributing the LDCs evenly on all the 3COM switches we demonstrated how the Summit 7i switch could effectively absorb the generated traffic without problems (top line in the graph). We then increased the outgoing traffic local to one single 3COM 4900 switch (we did this by adding more LDCs to a single switch, rather than distributing them across multiple switches). We had expected a saturation throughput of O(300) MB/s – the throughput equivalent to about three LDCs writing onto a triple Gigabit Ethernet trunk. We have instead observed only two thirds of that, a throughput of 220 Megabytes per second. Therefore, we knew we could use an outgoing capacity per 3COM switch equivalent to “only” two Gigabit Ethernet links. The incoming capacity of the 3COM switches and their triple Gigabit Ethernet trunk – on the other hand – reached the expected values and was never a problem.

We therefore had to take into account this “outgoing traffic” limitation for all the 3COM switches. To make things more difficult, we had also to make provisions for the outgoing traffic being written by DATE into itself and by DATE into the Permanent Data Storage. With the complete chain in place, we had to plan for a double load on all the outgoing links from the 3COM switches (half of the traffic created by the Data Acquisition system and half of the traffic created by the streaming from CASTOR to the Permanent Data Storage). We therefore had to re-define the topology of the whole test setup according to these findings.

Another problem came from the setup and operation of the computer farm. First of all, as all the machines were not available at the same time, the installation had to be done in stages. The Operating System, the framework, the products and all the annexed facilities had also to be re-installed on several nodes (due to updates, uniform deployment procedure and publication of new features). To make things more complicated, we have observed several dead on arrival (about 10% of the hosts), some failed-on-installation (another 25% of the hosts) and a few dead-on-operation machines. Hosts’ installation, restart and topological re-distribution took a considerable amount of manpower and we have never been able to use a system up to its planned run-time capacity. It is also true that the installation procedure took very little resources and was almost totally automated. The standard CERN Linux installation scheme, run jointly with a special script to install and configure the extra resources required by the ALICE Data Challenge, made a full reload of a node a routine procedure.

For the ALICE Data Challenge IV we made plans to use the new generation of StorageTek (STK) linear magnetic tape drives model 9940B, whose cartridges can accept up to 200 Gigabytes of data at a sustained rate of 30 MB/s. The units were delivered quite late and could be included in the test chain only at the very last moment. Clearly, this reduced considerably our flexibility for the deployment of the test and imposed hard constraints on the production period.

During the setup phase we have also noticed an issue arising from an incompatibility between the ALICE Mock Data Challenge output stream and the CASTOR input streams. Throughputs were very poor and the hosts’ resources (hardware and software) were clearly badly used. We later found out (unfortunately too late for the ALICE Data Challenge) that this was due to an architectural mismatch between the two components. We were therefore obliged to exclude the ALIMDC process from the ALICE Data Challenge chain and to replace it with a simpler front-end to the CASTOR system, writing raw DATE events into it. This proved to amply satisfy our requirements in time for the production period of the ALICE Data Challenge IV.

Several tests were performed on CASTOR. Some tests were run in isolation while other tests included a complete chain. The system behaved well if the incoming data remained below the maximum throughput that could be accepted by the tape devices. If instead the incoming throughput would exceed this value, CASTOR performances would degrade considerably, well below the maximum expected value.

During the planning phase for the ALICE Data Challenge we expected to integrate in the test setup new network technologies, namely some 10-Gigabit Ethernet equipment CERN had received in evaluation. This was done during the tuning stage of the final production period. The transition between the Gigabit Ethernet and the 10-Gigabit switches went almost transparently (we have seen some small troubles with the NFS distribution of the images and of the configuration files) and – at first – the results looked promising. Unfortunately we quickly encountered serious problems – total unrecoverable freeze of parts of the network – that forced us to switch back to the original setup based on Gigabit Ethernet. The problem was later identified as an issue related to the integration of a special ASIC used to handle the communication inside the 10 Gigabit Ethernet switches. The manufacturer issued a fix, unfortunately too late to reintegrate the now (apparently) working material in the ALICE Data Challenge setup.

## 3.8. Production periods

During the ALICE Data Challenge IV, milestones were distributed over two production periods.

The first production period was held in July 2002, when high-rate raw data was transferred within DATE over a relative short period, target being 650 MB/s sent from the event builder to the null device.

The second period has the objective to achieve 200 Megabytes per second sustained to tape for a minimum of 7 consecutive days and to create at the same time a data set of about 200 TB of data in Permanent Data Storage (PDS). For this milestone the complete chain (LDCs to GDCs to CASTOR to tapes) had to be active. For this reason, we had to wait for the delivery of the required tape units (this happened in late November 2002), to go through a successive validation period and – finally – to coordinate the allocation of a considerable amount of CERN public resources. The second test session was started on December 6$^{th}$, 2002.

## 3.9. Outcomes

The ALICE Data Challenge IV can be considered as a complete success. As we will see, both milestones have been met – if not exceeded – using production-like software and standard tools and services.

The first milestone – sustained throughput of 650 MB/s through DATE – was met on July 2nd 2002, when event building reached the aggregate throughput of 1.8 Gigabytes per second.

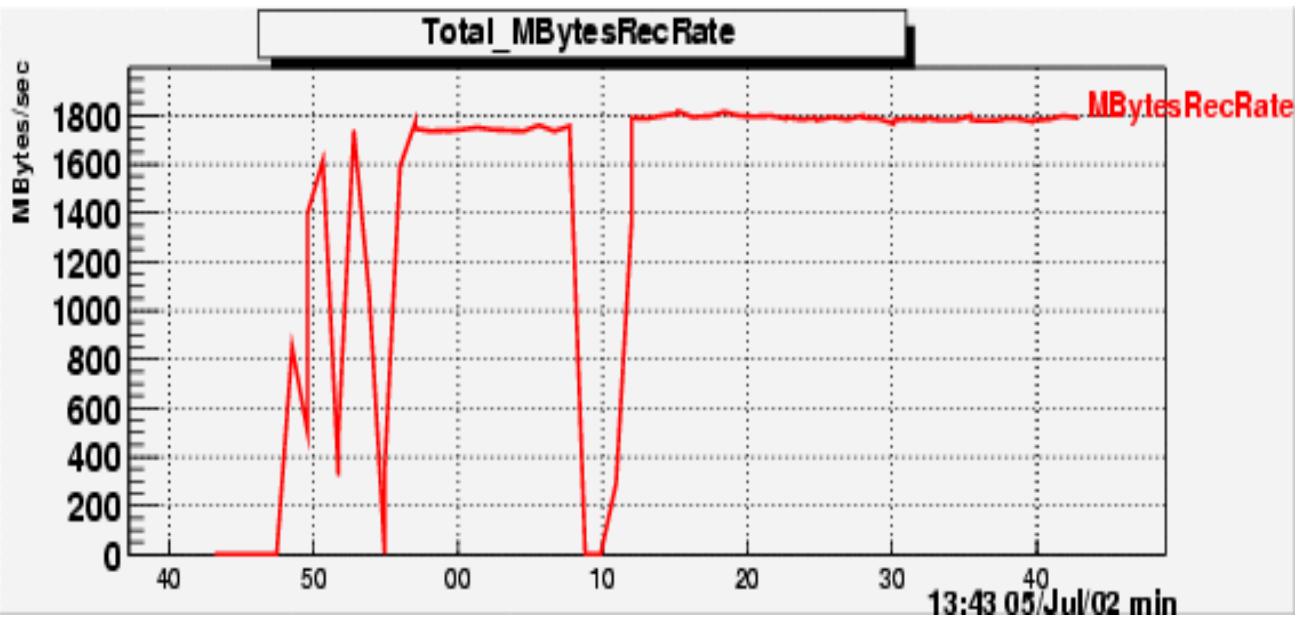

Figure 9: DATE raw event building throughput.

The graph reported in Figure 9 shows the monitoring information published by AFFAIR on a Data Acquisition system composed of 40 LDCs and 38 GDCs writing fixed size and fixed pattern events of 40 Megabytes each (1 Megabyte per LDC) to the null device. Incidentally, the 1.8 GB/s is also the theoretical limit imposed by the output trunks from the eight 3COM switches (~220 MB/s per trunk distributed on 8 trunks). The above throughput is therefore the theoretical maximum that we could have squeezed out of the deployed network topology. The milestone was quickly achieved and no major problems were encountered. The only issue behind this milestone was the unavailability of several machines (dead on arrival, failed on installation, failed on operation) that required a careful and detailed optimization of the available resources. Little Operating System tuning was needed and on the LDCs and the GDCs we had plenty of spare system resources available. The LDCs had about ¼ of one CPU free and the GDCs had about 1 CPU free. The test was run on the LCG testbed, with hosts based on dual-Pentium III CPUs.

The second milestone required more detailed setup and careful tuning. The only fact that we had to make intensive use of public CERN resources (network backbone, tape robots, tape units, tape libraries, servers) imposed hard constraints on the schedule of the various test phases. It was also the first time that CASTOR v1.4.1.7 was attached to a stream carrying such a bandwith. If we add the fact that several of the hardware components had never been used before on a system of this scale, we clearly might have had the perfect recipe for a disaster. This was not the case: all components behaved as expected and we had – at least at first – very little problems to get things going. Previous tests demonstrated how the deployed PDS setup could not accept more than a given amount of data and we therefore limited ourselves to this amount (well above our planning requirements). We also opted for a "relaxed" operator intervention policy, limited to working hours and to a few occasional checks after hours or during the weekend. With all this is mind, the system performed as expected, even recovering from some degradations introduced by the failure of one of the tape units (that had later to be removed from the test setup) and by some reconfigurations that followed. The measures made with AFFAIR over the test period are shown in Figure 10.

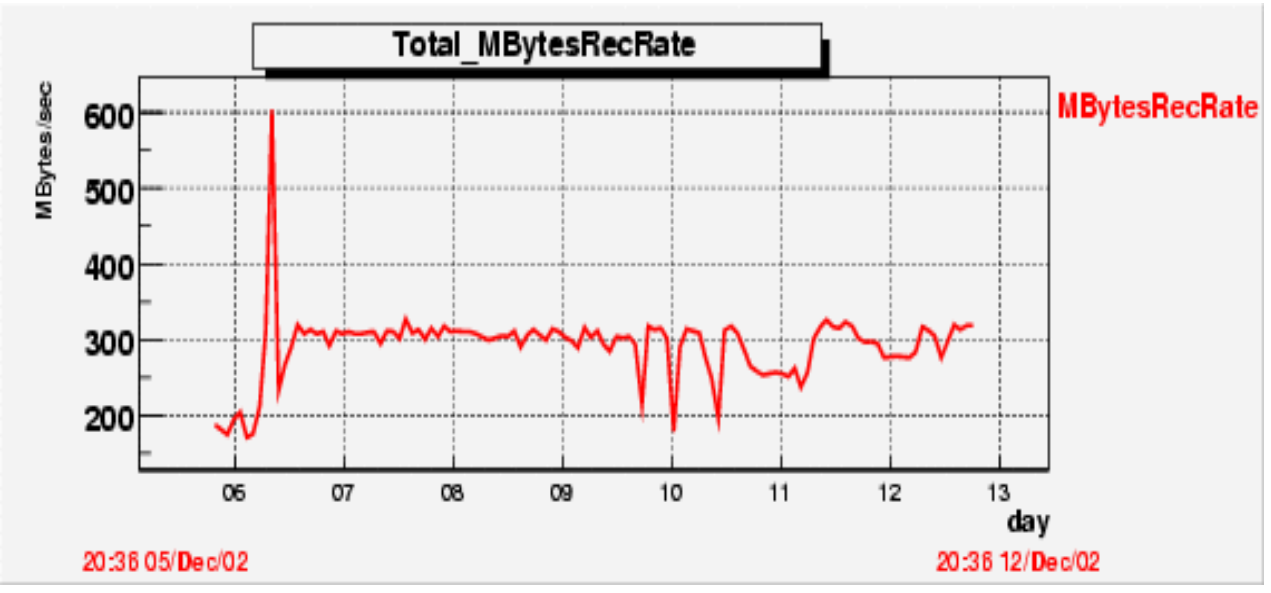

Figure 10: sustained throughput milestone monitoring.

After a sharp ramp-up period on the December 6th – when not-yet-full disks could accept data at nominal bandwith – the system behaved well for four days, when – after a first warning sign in the evening of the 9th – one tape unit failed and the whole system had to be reconfigured. Following this phase, the throughput returned to the nominal value on December 11th to remain stable until the end of the test. The final results were: a peak rate of 310 Megabytes per second (ramp-up period excluded), a sustained rate of 280 Megabytes per second and 180 Terabytes moved onto Permanent Data Storage for a time period of seven consecutive days.

## 4. FUTURE DATA CHALLENGES

We feel that several important issues have not been adequately confronted during the ALICE Data Challenge IV. They have since been reviewed and will play a role in the planning of the future ALICE Data Challenges.

ALICE-like data pattern must be correctly deployed. ALICE will not move data streams of complex structure and this will be an important factor for all future tests and production periods. This may imply the use of different network topologies, the deployment of new network technologies (NICs and switches) and the allocation of dedicated tuning and setup periods in our program of work.

Online handling of the data coming from the LDC emulators must be tried out. This shall include the objectification of raw data events and some on-the-fly reconstruction processing. In the ALICE Data Challenge IV these objectives had to be dropped due to time constraints and lack of resources. Furthermore, data analysis implies a certain structure and format of the data, to be agreed between the ALICE Online and Offline teams. Good progresses have been made on this issue and we are confident for future Challenges, when we expect ROOT objects to be stored and distributed to some selected Tiers outside CERN. The ALICE Environment – AliEn – [6] is ready for integration to the ALICE Data Challenge setup and shall be soon tested in "real life" conditions.

So far we have always emulated the data stream created from the ALICE detectors via a software module. Since some time now, a hardware data source emulator is available from the ALICE Data Acquisition group. For the future data challenge we expect to integrate at least one complete chain at the input of the data streams and to feed this into the raw data path. This will be an important milestone, as we will have – for the first time – the ALICE readout card (the pRORC) part of an ALICE-like Data Acquisition system. We expect to setup the Detector Data Link (DDL) chain in the ALICE DAQ lab, directly linked by the CERN backbone to the LCG test setup via a Gigabit Ethernet uplink.

The 300 MB/s barrier to PDS observed in 2002 will have to be broken. We have already reached the milestone planned for the ALICE Data Challenge year 2003 (300 Megabytes per second). However, the LCG testbed plans an upgrade to 450 Megabytes per second for the year 2003 and we shall profit from this extra bandwith. We know that the Data Acquisition system is capable of throughputs much higher than that, so we have good hopes in what the forthcoming host computers, network and tape technologies will be able to give to us.

## 5. CONCLUSIONS

The ALICE Data Challenge IV proved to be a valuable input for future developments as well as a successful exercise to achieve very important – for CERN and for the ALICE collaboration – milestones. A rather significant set of equipment was put together to form an ALICE-like Data Acquisition setup. The output data stream was successfully recorded onto Permanent Data Storage with excellent rates, reliability and stability. Commercial components and CERN in-house packages integrated at the best of expectations. All the milestones were met – several even exceeded – and we are now between one and three years ahead of the proposed planning. This does not mean that we are out of work, on the contrary. The ALICE collaboration has stringent and difficult requirements that will always justify the deployment of new, more demanding ALICE Data Challenges. New technologies, products, libraries and developments will require the preparation, setup and operation of similar exercises. Only in this way we will be able to guarantee a reasonable level of confidence in the complete data chain once the first events will be triggered at the LHC collider: by challenging the challenge.